\documentclass[lettersize,journal]{IEEEtran}
\usepackage{amsmath,amsfonts}
\usepackage{algorithmic}
\usepackage{algorithm}
\usepackage{array}
\usepackage{tabularx}
\usepackage[caption=false,font=normalsize,labelfont=sf,textfont=sf]{subfig}
\usepackage{textcomp}
\usepackage{stfloats}
\usepackage{url}
\usepackage{verbatim}
\usepackage{graphicx}
\usepackage{cite}
\usepackage{multirow}
\usepackage{booktabs}
\usepackage{xcolor}
\usepackage{xcolor}
\begin{document}

\title{Methodological Pitfalls in Predicting Ship Turnaround Time at Brazilian Ports: An Empirical Audit and a Reproducible Pipeline}

\author{Eduardo Caruso Barbosa Pacheco, Reynaldo Pereira Martins, Alexandre dos Santos Gualberto,  Jo\~ao Vitor Souza Germano, Milton Miranda Neto and Francisco Louzada
\thanks{The authors are with the Institute of Mathematical and Computer Sciences (ICMC), University of S\~ao Paulo, S\~ao Carlos, Brazil (e-mail: eduardo.caruso.pacheco@usp.br; reynaldo.martins@usp.br; alexandre.gualberto@usp.br; joao.germano@usp.br; miltonmirandaneto@usp.br; louzada@icmc.usp.br)).}}

\markboth{IEEE Transactions on Intelligent Transportation Systems,~Vol.~XX, No.~X, 2026}%
{Pacheco \MakeLowercase{\textit{et al.}}: Methodological Pitfalls in Predicting Ship Turnaround Time at Brazilian Ports}


\maketitle

\begin{abstract}
Two recent studies on ship turnaround time at Brazilian ports report models with R-squared scores above 0.80 on data from the national waterway transportation agency. We audit their published datasets and reveal these results are inflated by structural data leakage. Specifically, 97 percent of berthing events in their random row-level split appear in both training and testing partitions. We demonstrate that a model's R-squared drops from 0.82 to 0.20 when the data is properly split at the berthing level. We further document additional methodological flaws, including survivorship bias, undocumented target truncation, and the retention of variables that algebraically reconstruct the target. To address these issues, we propose a reproducible pipeline that aggregates cargo to the berthing level, utilizes conditional imputation, and represents high-cardinality cargo variables through weighted averages of multilingual sentence embeddings. Evaluating six time targets and thirty seeds, our pipeline reveals that cargo embeddings yield statistically significant improvements on four operational targets, while administrative waiting times remain unaffected. Prior works missed this per-target heterogeneity by conflating genuine signal with leaked identity.
\end{abstract}


\section{Introduction}
\IEEEPARstart{M}{aritime} transport carries roughly 80\% of international trade by volume, and bottlenecks in ship turnaround time at ports propagate through the entire logistics chain. The Brazilian National Waterway Transportation Agency (ANTAQ) has published operational records of berthings and cargo since 2010, and two recent studies have applied machine learning to these data with promising headline numbers: Abreu et al.~\cite{abreu2023} report a Random~Forest classifier with accuracy and $F_1$ of 73.9\% on a four-class discretization of the stay time, and Rao et al.~\cite{rao2025} report a Random~Forest regressor with $R^2 = 0.829$ and $\mathrm{RMSE} = 10.44$~h on the same dataset. These numbers are good enough to be surprising. We obtained the datasets the authors published, ran the operations they describe, and found that their numbers are consistent with a pipeline that systematically inflates accuracy through structural leakage, undocumented truncation of the target distribution, and silent inclusion of variables that algebraically reconstruct the target. None of these problems is a coding bug. They are all reasonable-looking choices that produce the wrong number when combined with the one-to-many shape of the source database.

This article does two things. First, it audits empirically what these published pipelines do, with code and numbers. We are not just claiming there is leakage; we measure how much (97\% of test-set berthings appear in the training set), and we demonstrate that a one-line lookup table beats their best model under their split (Section~\ref{sec:critique}). Second, we propose a corrected pipeline---aggregation, conditional imputation, semantic embeddings, and a defensible outlier filter---and report what realistic predictive performance on this problem looks like (Sections~\ref{sec:pipeline} to~\ref{sec:results}). Across 30 seeds, 6 time targets and 4 configurations we find substantial and statistically significant gains from semantic cargo representations on operational time targets (Cohen's $d$ from $3.8$ to $17.3$), and no measurable gain on administrative waiting targets. This last finding---that the impact of cargo information is target-dependent---is something that the published pipelines could not have observed because the leakage drowns the signal.

\textbf{Contributions.}
\begin{itemize}
    \item An empirical audit, with executable code, of the data and pipeline published by Abreu et al.\ and Rao et al., quantifying leakage at the row level and identifying four additional issues that were not previously catalogued.
    \item A reproducible processing pipeline that uses one-to-many aggregation, port-conditional imputation, multilingual sentence embeddings weighted by cargo mass, and a semantic outlier filter.
    \item A controlled ablation over 30 seeds, 6 targets and 4 configurations with paired Wilcoxon tests and Cohen's $d$, showing heterogeneous, target-specific effects of cargo representation.
    \item All code, the cleaned dataset (2018--2024), trained models and seed-by-seed metrics are released to support reproduction.
\end{itemize}
\subsection{Aggregation at the Berthing Level}

A preliminary conceptual discussion of the methodological issues surrounding data leakage and survivorship bias in this domain was presented as an extended abstract at the 4th Workshop on Applied Mathematics, Statistics, and Computing to Industry (WMECAI)~\cite{pacheco2025}. This journal submission substantially extends that prior work. While the conference presentation merely theorized potential solutions, this paper fully implements and evaluates a reproducible pipeline over a vastly expanded seven-year dataset (2018–2024). Furthermore, this manuscript introduces rigorous empirical contributions not present in the preliminary work, including a quantified audit of the original models with executable code, the expansion from a single target to six distinct operational and administrative time targets, and a controlled ablation study over 30 random seeds to demonstrate the statistically significant impact of multilingual semantic cargo embeddings.

To support the reproducibility of our findings, all code and pre-trained models are publicly available at https://github.com/eduardocbpacheco/ANTAQ . All datasets are available at Zenodo\cite{antaqdata2026} 
\section{Related Work}
\label{sec:related}

\textbf{Abreu et al.~(2023)}~\cite{abreu2023} train thirteen classifiers on ANTAQ 2018 data (the most recent year available to them at the time, a limitation they explicitly acknowledge in their conclusions) to predict \texttt{TStay} discretized into four equal-width classes. After a data-cleaning step that removes columns with missing values, constant columns, multicollinearity above 0.90, high cardinality and skewness, and after an SVM-RBF outlier removal, 150{,}669 rows remain. Random~Forest is the best model, with 73.9\% accuracy and $F_1$. The top features by information gain are geographical (port, city, port complex, region) and operational (cargo operation type, inland route).

\textbf{Rao et al.~(2025)}~\cite{rao2025} take the same dataset (which they describe as covering ``11/2017--12/2018,'' although the released CSV contains only 2018 events) and frame both regression and four-class classification of total time, and binary classification of delay time. They explicitly state ``the data ($n = 153{,}331$) are split into 80\% (122{,}664) for training and 20\% (30{,}667) for testing'' and use target encoding for high-cardinality variables including \texttt{Berth}, \texttt{Source}, \texttt{Destination}, and \texttt{Port}. Their Random~Forest reports $R^2 = 0.829$, $\mathrm{RMSE} = 10.44$~h for total time, and \texttt{Berth} appears as the most important feature in their SHAP plot. Both papers describe their pipelines in enough detail that they can be audited against the published data. The next section formalizes the prediction problem we are studying, and the section after audits the data.

Vessel-side ETA and turnaround-time prediction is an active area internationally. Jahangard et al.~\cite{jahangard2025} survey 124 papers on machine learning and optimization for seaport operations and index the Brazilian-port stay-time results among the predictive entries without independent audit. Recent primary studies on adjacent ports include Chu et al.~\cite{chu2024} on vessel turnaround time at Hong Kong (XGBoost~\cite{chen2016} with a temporal split and the shipper-reported estimated departure time as a feature), Pham and Nguyen~\cite{pham2025} on ship berthing time at Tan Cang Cat Lai, Park et al.~\cite{park2026} on Busan terminal time-series with a GRU benchmarked against six reimplemented competitors, Gülmez et al.~\cite{gulmez2024} on cargo handling and berthing times in a bulk terminal, and Saber et al.~\cite{saber2025} on ETA in the Baltic from AIS trajectories. All of these works cite Abreu et al.\ or Rao et al.\ as part of the literature on machine learning for stay-time prediction, and none examines the ANTAQ source data or the one-to-many cardinality of berthing-to-cargo records. The headline $R^2 \approx 0.83$ from the Brazilian-port literature thus functions as an external anchor in the wider field: to the extent that it is inflated by the structural leakage we document in Section~\ref{sec:critique}, the inflation propagates whenever the figure is cited as evidence of what is achievable on this class of problem.

\section{Problem Formulation}
\label{sec:problem}

Let $\mathcal{B}$ denote the set of berthing events at a Brazilian port, where each event $b \in \mathcal{B}$ is a single visit of a ship to a berth with arrival, mooring and unmooring timestamps. The supervised learning problem we consider is to predict, for each berthing, six target durations $y^{(t)}_b \in \mathbb{R}_{\geq 0}$ with $t \in \{\text{Stay},\text{WaitMooring},\text{Moored},\text{WaitStartOp},\text{Operation},\text{WaitUnmooring}\}$, from a feature vector $x_b \in \mathbb{R}^p$ describing the operating context. The total stay decomposes as $y^{(\text{Stay})}_b = y^{(\text{WaitMooring})}_b + y^{(\text{Moored})}_b$, with $y^{(\text{Moored})}_b = y^{(\text{WaitStartOp})}_b + y^{(\text{Operation})}_b + y^{(\text{WaitUnmooring})}_b$, so the targets are not algebraically independent. We treat them as separate regression problems on the same matrix.

Given a dataset $\mathcal{D} = \{(x_b, y^{(t)}_b)\}_{b \in \mathcal{B}}$ and a class of predictors $f(\,\cdot\,; \theta)$ with parameters $\theta$, the standard empirical risk minimization (ERM) objective is
\begin{equation}
\hat{\theta} = \arg\min_\theta \frac{1}{|\mathcal{B}|}\sum_{b \in \mathcal{B}} \big(y^{(t)}_b - f(x_b; \theta)\big)^2,
\label{eq:erm}
\end{equation}
and the standard generalization argument requires that the events $b$ in $\mathcal{D}$ be approximately i.i.d.\ realizations of the data-generating process. This is the assumption that the prior literature violates structurally, not by misimplementation of any one step but by choosing $b = \text{cargo row}$ in (\ref{eq:erm}) instead of $b = \text{berthing}$. The ANTAQ source database stores berthings (one row per ship visit, one row per target value) and cargo records (one row per load moved, many rows per berthing) in two related tables. The natural pseudo-i.i.d.\ unit for prediction is the berthing: each berthing is one visit of a ship, with one set of times, occurring once. Joining the cargo table into the berthing table without aggregation transforms the unit of analysis into the cargo row, which is not i.i.d.\ with respect to the targets---multiple cargo rows from the same berthing carry identical $y^{(t)}_b$. Under a random split of cargo rows, the same berthing appears in both partitions, the i.i.d.\ assumption of (\ref{eq:erm}) is violated, and the test risk
\[
\hat{R}_{\text{test}}(\hat{\theta}) = \frac{1}{|\mathcal{B}_{\text{test}}|}\sum_b \big(y_b - f(x_b; \hat\theta)\big)^2
\]
underestimates the population risk that the model would face on unseen berthings. This is the formal version of the leakage we audit empirically in Section~\ref{sec:critique}; it is a textbook case of \emph{target leakage} as classified by Kaufman et al.~\cite{kaufman2012}, and a special case of the broader replicability failures catalogued in recent surveys of applied ML~\cite{kapoor2023, lones2024}.

\section{The ANTAQ Dataset}
\label{sec:data}

The audit in Section~\ref{sec:critique} operates on the two pre-joined CSV files released by Abreu et al. Our own pipeline instead starts from the primary ANTAQ open microdata (\emph{Estat\'istico Aquavi\'ario}), which the agency publishes once per year as a set of semicolon-delimited text files organized as a star schema around two fact tables. We download the seven years 2018--2024 directly from the agency and process them with the pipeline of Section~\ref{sec:pipeline}. This section describes the source tables, the prediction targets, and the raw fields we select as features, so that every later transformation can be traced back to a documented input.

Table~\ref{tab:source_tables} lists the tables we use and their grain. The two fact tables are \texttt{Atracacao} (berthing) and \texttt{Carga} (cargo): a single berthing---one ship visit to a berth---generates one \texttt{Atracacao} row and, typically, many \texttt{Carga} rows, one per load moved during the visit. In 2018 alone the agency files contain 76{,}441 berthings against roughly 2.09 million cargo rows---the same one-to-many fan-out that drives the leakage of Section~\ref{sec:critique}, here at the full granularity of the source rather than in the truncated published snapshot. The remaining tables are satellites: \texttt{TemposAtracacao} carries the six duration targets keyed by berthing; \texttt{Carga\_Conteinerizada} holds the specific containerized commodity and its weight; \texttt{Carga\_Rio}, \texttt{Carga\_Hidrovia} and \texttt{Carga\_Regiao} hold the financial value moved on each river, waterway and hydrographic region; and four dimension tables (\texttt{Mercadoria}, \texttt{MercadoriaConteinerizada}, \texttt{Instalacao\_Origem}, \texttt{Instalacao\_Destino}) resolve numeric codes into the Portuguese text labels we later embed.
\begin{table}[htbp]
\centering
\caption{The six prediction targets, taken from the \texttt{TemposAtracacao} table (hours). We use the short names in the first column throughout the paper.}
\label{tab:targets}
\small
\setlength{\tabcolsep}{4pt} 
\begin{tabular}{@{} p{0.45\columnwidth} p{0.50\columnwidth} @{}}
\toprule
\textbf{Target (ANTAQ name)} & \textbf{Interval it measures} \\ \midrule
\texttt{TStay} \newline (\texttt{TEstadia}) & arrival $\to$ unmooring (total turnaround) \\
\texttt{TWaitMooring} \newline (\texttt{TEsperaAtracacao}) & arrival $\to$ mooring (queue / anchorage wait) \\
\texttt{TMoored} \newline (\texttt{TAtracado}) & mooring $\to$ unmooring \\
\texttt{TWaitStartOp} \newline (\texttt{TEsperaInicioOp}) & mooring $\to$ operation start (administrative) \\
\texttt{TOperation} \newline (\texttt{TOperacao}) & operation start $\to$ operation end (loading / unloading) \\
\texttt{TWaitUnmooring} \newline (\texttt{TEsperaDesatracacao}) & operation end $\to$ unmooring (administrative) \\
\bottomrule
\end{tabular}
\end{table}
\begin{table*}[!t]
\centering
\caption{ANTAQ source tables used by the pipeline. Row counts are for
2018; the same structure repeats for each year 2018--2024.}
\label{tab:source_tables}

\footnotesize
\setlength{\tabcolsep}{3pt}
\renewcommand{\arraystretch}{1.12}

\begin{tabularx}{\textwidth}{@{}
>{\raggedright\arraybackslash}p{0.31\textwidth}
>{\raggedright\arraybackslash}p{0.25\textwidth}
>{\raggedright\arraybackslash}X
@{}}

\toprule
\textbf{Table}
&
\textbf{Grain (one row per\ldots)}
&
\textbf{Key contents}
\\
\midrule

\texttt{Atracacao} (Berthing)
&
berthing (76{,}441)
&
berthing, berth, port and complex identifiers; municipality, state,
geographic and hydrographic region; arrival, mooring, unmooring and
operation start/end timestamps; operation and navigation type;
shipowner nationality
\\

\texttt{TemposAtracacao}
&
berthing (76{,}441)
&
the six duration targets of Table~\ref{tab:targets}
\\

\texttt{Carga} (Cargo)
&
cargo load moved ($\approx 2.09$\,M)
&
gross weight, TEU, count, origin, destination, cargo nature, cargo
operation type, direction, container status, navigation and cabotage
flags
\\

\texttt{Carga\_Conteinerizada}
&
containerized item
&
containerized merchandise code and weight
\\

\texttt{Carga\_Rio} / \texttt{\_Hidrovia} /
\texttt{\_Regiao}
&
cargo $\times$ river / waterway / hydrographic region
&
financial value moved on each
\\

\texttt{Mercadoria}
\newline
\texttt{MercadoriaConteinerizada}
&
merchandise code
&
Portuguese merchandise-group text label
\\

\texttt{Instalacao\_Origem}
\newline
\texttt{Instalacao\_Destino}
&
facility code
&
origin / destination facility text label and coordinates
\\

\bottomrule
\end{tabularx}
\end{table*}

The \texttt{TemposAtracacao} table provides six non-negative durations per berthing, all in hours, which are our regression targets (Table~\ref{tab:targets}). Throughout the paper we refer to them by the short English names in the first column. They are not independent: the total stay decomposes as $\texttt{TStay} = \texttt{TWaitMooring} + \texttt{TMoored}$ and the moored interval as $\texttt{TMoored} = \texttt{TWaitStartOp} + \texttt{TOperation} + \texttt{TWaitUnmooring}$, as formalized in Section~\ref{sec:problem}. We model each as a separate regression on the same feature matrix rather than predicting one and deriving the rest, because the per-target sensitivity to cargo turns out to be heterogeneous (Section~\ref{sec:results}): the two administrative waits (\texttt{TWaitStartOp}, \texttt{TWaitUnmooring}) behave very differently from the operational intervals, a structure that the single aggregate target of the prior work cannot reveal.

\label{sec:feature_selection}
From the fields above we keep only those available to a shipper before the vessel arrives and discard the rest as leakage (Section~\ref{sec:leak_decisions}). The retained fields fall into five roles. (i) \emph{Berthing-level categoricals} that are constant within a berthing by construction---mooring port, port complex, port-authority type, operation type, navigation type, shipowner nationality, municipality, state (UF), geographic region, hydrographic region, and the river-installation flag---are carried unchanged. (ii) Four \emph{cargo categoricals that are near-homogeneous} within a berthing (cargo nature, inland-transport route, cargo-nature status, and general-cargo packaging, each varying in under 3\% of berthings) are reduced with \texttt{FIRST()}. (iii) Nine \emph{heterogeneous cargo categoricals} (cargo operation type, direction, container status, container-size flag, and the long-course, cabotage and inland-transport flags) vary within 5--18\% of berthings and are expanded into the weighted one-hot-sum of Section~\ref{sec:ohe_sum}. (iv) \emph{Cargo numerics} (gross weight, TEU, container count, containerized weight, and the four financial values) are summed to the berthing level. (v) The four \emph{high-cardinality text variables}---cargo group, containerized cargo group, origin and destination, with 78, 98, 342 and 466 distinct values---are represented by the multilingual embeddings of Section~\ref{sec:embeddings}. To these we add the port longitude and latitude and the cyclic encoding of the month (Section~\ref{sec:pipeline}). Berth, the absolute event dates, and the five partial times are deliberately excluded as model inputs for the reasons given in Section~\ref{sec:critique}; the partial times are retained only as the alternative regression targets of Table~\ref{tab:targets}. The resulting column counts per configuration are summarized later in Table~\ref{tab:feature_breakdown}.
\section{Empirical Audit of Prior Work}
\label{sec:critique}

We obtained the two CSV files that Abreu et al.\ released with their paper. The first, \texttt{CargasBR2018EDA-LOGII.csv} (153{,}331 rows $\times$ 35 columns, here ``EDA file'' for Exploratory Data Analysis), is the result of joining the ANTAQ \emph{Berthing} table with the \emph{Cargo} table by berthing identifier without aggregation; it keeps the cargo origin and destination, the absolute event dates, and the five partial-time variables (waits and operation time). It is the dataset Rao et al.\ use in their experiments. The second, \texttt{EN-CargasBR2018Modelo-LOGII.csv} (150{,}669 rows $\times$ 21 columns, here ``Model file''), is the post-cleaning version that Abreu et al.\ feed to the Random~Forest: high-cardinality fields (\texttt{Origin}, \texttt{Destination}), absolute dates, and the partial-time variables have been dropped, and the regression target has been replaced by the four-class \texttt{ClassStay}. We ran the operations both papers describe on these two files and measured what happens.

\subsection{Structural Leakage from One-to-Many Joins}

\begin{figure}[!t]
\centering
\includegraphics[width=\columnwidth]{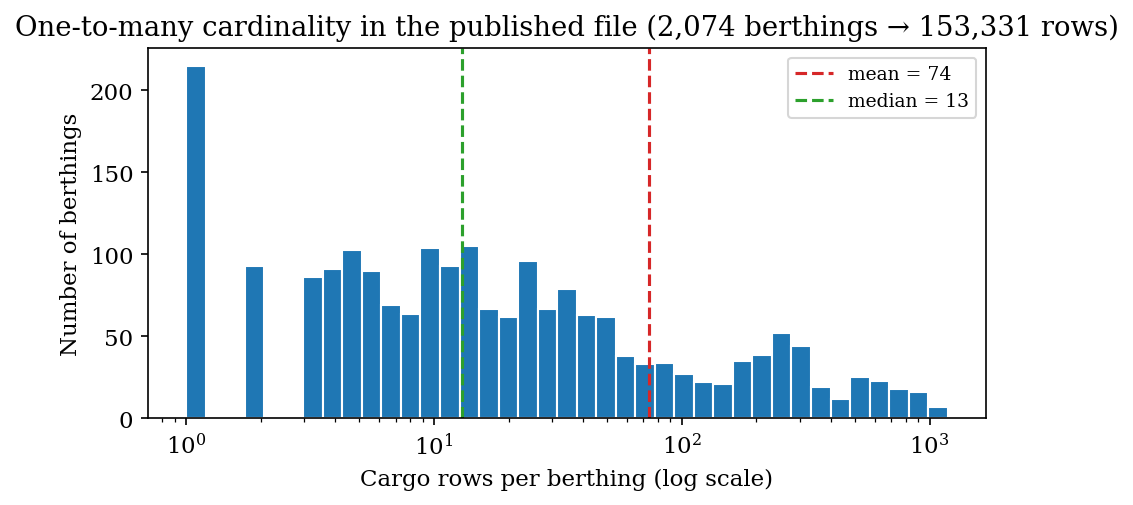}
\caption{One-to-many cardinality between berthings and cargo rows in the EDA file published by Abreu et al. Histogram of the number of cargo rows per berthing in log scale.}
\label{fig:cardinality}
\end{figure}

The EDA file is the result of joining berthing records with cargo records by \texttt{IDBerthing}. A single berthing typically corresponds to many cargo lines: in the published dataset, 2{,}074 unique berthings produce 153{,}331 rows --- an average of 74 cargo rows per berthing, median 13, maximum 1{,}183 (Fig.~\ref{fig:cardinality}). The stay time is, by definition, constant within a berthing: the standard deviation of \texttt{TStay} within a berthing is exactly zero for 2{,}073 of the 2{,}074 berthings.

A random 80/20 row-level split, as both papers use, therefore puts about 97\% of the test-set berthings also in the training set, each with the same target value. Our reproduction of the published split places 1{,}597 of the 1{,}644 distinct berthings of the test set within training set. The model doesn't need to predict; it only has to memorize the berthing identity.

To quantify the leakage, we ran three experiments on the same published CSV (Table~\ref{tab:leakage_levels}). The first is a feature-free lookup that returns the mean \texttt{TStay} of the training rows that share the (Mooring Port, Berth, Mooring Date) identity; on a random row-level split it reaches $R^2 = 0.9950$ and $\mathrm{RMSE} = 1.77$~h, and on a berthing-level split it collapses to $R^2 \approx 0$. This lookup measures the \emph{ceiling} of what is recoverable from the leaked identity when the full key is available; it is not the limit of the Rao et al.\ model because Rao et al.\ deliberately drop absolute dates and keep only \texttt{Month}, so their Random~Forest cannot reach that ceiling. The second experiment is a Random~Forest regressor (the same model class as Rao et al.) with the same feature set: target encoding of \texttt{Berth}, \texttt{Source}, \texttt{Destination}, \texttt{Port}, no absolute dates, default sklearn hyperparameters (\texttt{n\_estimators=100}, unbounded depth), 25 random seeds. Under the row-level split it reaches $R^2 = 0.821 \pm 0.005$, essentially the same as the $R^2 = 0.829$ reported by Rao et al. The third experiment is the same Random~Forest on the same 25 seeds, but under a berthing-level split: $R^2$ drops to $0.204 \pm 0.144$. The mean paired difference is $\Delta R^2 = 0.617 \pm 0.146$ with Cohen's $d = 4.24$ and Wilcoxon $p < 3 \times 10^{-8}$. The same model on the same data loses roughly $0.62$ of $R^2$ when the split is moved from the row level to the berthing level---this is the portion of the reported performance attributable to leakage in the form that the original models can access.

\begin{table}[!t]
\centering
\caption{R\textsuperscript{2} on the same published dataset under different identifying conditions. The Random~Forest rows aggregate 25 random seeds with default sklearn hyperparameters, matching the configuration of Rao et al. The paired comparison between the two Random~Forest rows yields Cohen's $d = 4.24$ at $p < 3 \times 10^{-8}$.}
\label{tab:leakage_levels}
\small
\setlength{\tabcolsep}{4pt}
\begin{tabular}{@{}p{0.34\columnwidth}ccc@{}}
\toprule
\textbf{Experiment} & \textbf{Identity} & \textbf{Split} & $R^2$ \\ \midrule
Lookup by (Port, Berth, Date) & full key & row & \textbf{0.995} \\
Random~Forest (Rao-like features) & partial & row & \textbf{0.821 $\pm$ 0.005} \\
Rao et al.\ Random~Forest (reported) & partial & row & 0.829 \\
Same Random~Forest & partial & berthing & \textbf{0.204 $\pm$ 0.144} \\
\bottomrule
\end{tabular}
\end{table}

\subsection{Survivorship Bias and Implausible Targets}

The transition from the EDA file (153,331 rows) to the Model file (150,669 rows) involves the removal of 2,662 rows, representing a 1.7\% reduction. The authors attribute this to the deletion of remaining missing values after column-level cleaning, but neither paper publishes the exact rule. In maritime datasets, missingness is rarely random; discarding rows rather than imputing them introduces survivorship bias by systematically silencing specific operational regimes that may systematically lack certain records.  More problematic is the retention of physically impossible data: 1,027 rows in the EDA file record a TStay exactly equal to zero. Although a berthing event of zero duration is impossible, these rows survive the opaque cleaning process and heavily contaminate the low end of the distribution. This structural artifact artificially inflates the “Low” class in Abreu et al. and skews the lower bin in Rao et al., compromising their models' ability to learn genuine short-stay dynamics. 

\begin{figure*}[!t]
\centering
\includegraphics[width=\textwidth]{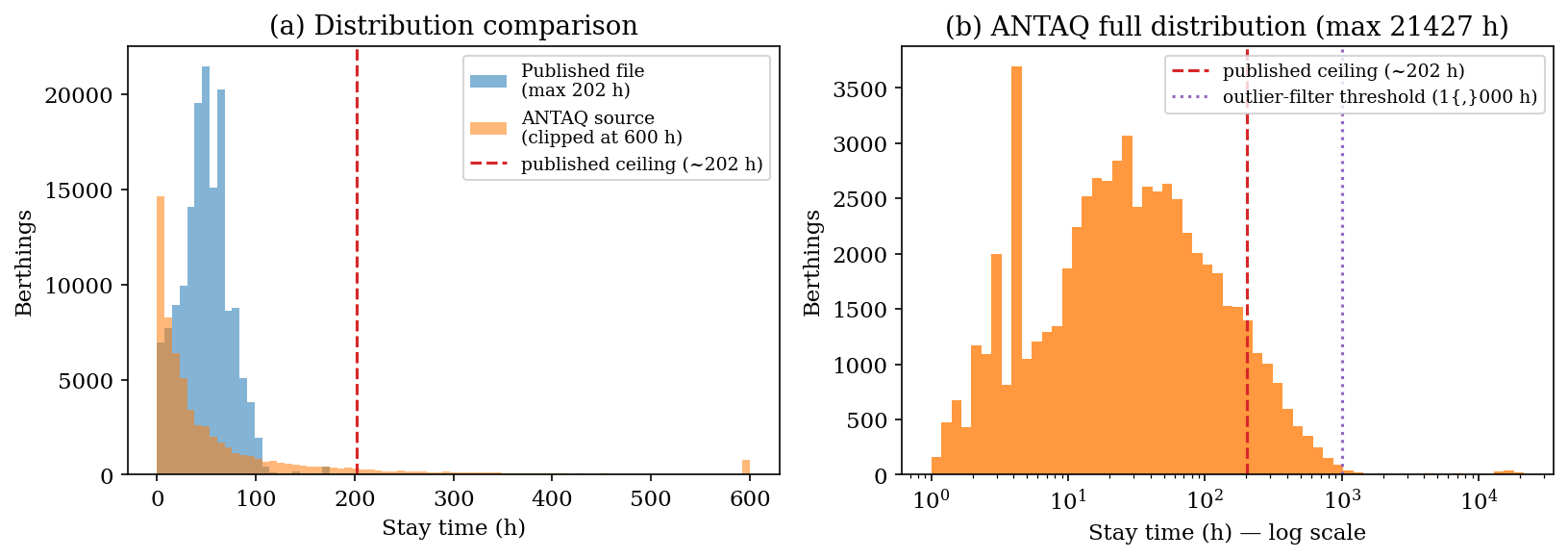}
\caption{Distribution of \texttt{TStay}. (a) The histogram of the EDA file (in blue) sits entirely below 202~h while the ANTAQ source distribution (in orange, clipped at 600~h for display) carries a long tail. (b) Same ANTAQ source data in log scale, showing the tail up to 21{,}427~h. The dashed red line marks the silent ceiling of the published file; the dotted purple line marks the threshold of our semantic outlier filter (Section~\ref{sec:pipeline}).}
\label{fig:ceiling}
\end{figure*}

\subsection{An Undocumented 202-Hour Ceiling}

The histograms of \texttt{TStay} in both papers stop at 200~h. The maximum value in the EDA file is 201.99~h. The full ANTAQ 2018 data we downloaded directly from the agency contains berthings up to 21{,}427~h (about 2.5 years), with 9.25\% of berthings above 200~h and 0.37\% above 1{,}000~h. The published file silently truncates this tail (Fig.~\ref{fig:ceiling}). Abreu et al.\ describe an SVM-RBF unsupervised outlier removal step in §5.2 of their paper, and the 202-h ceiling is consistent with that step having produced this cut, although neither paper publishes the threshold or the number of rows it removed. The net effect is that both papers report performance on a population that has been pre-filtered to look easier than it is, without a fully documented rule.

\subsection{Berth as Leakage at the Decision Point}

Rao et al.\ identify \texttt{Berth} as the most important feature in their Random~Forest, and reinforce this finding through SHAP analysis. However, the berth is assigned by the port at the moment of mooring; the shipper does not know it at the time the prediction would be useful. Under the decision point ``shipper, before arrival,'' berth is a future signal. We verified its predictive power directly: under a berthing-level split, a one-line lookup that uses only \texttt{Berth} reaches $R^2 = 0.32$ on the test set---substantial, but only available after the fact.

\subsection{Partial-Time Variables That Are Hard to Audit}

The EDA file retains the five partial times \texttt{TWaitMooring}, \texttt{TWaitStartOp}, \texttt{TOperation}, \texttt{TWaitUnmooring}, \texttt{TMoored}. We verified algebraically that, to numerical precision, $\texttt{TStay} = \texttt{TWaitMooring} + \texttt{TWaitStartOp} + \texttt{TOperation} + \texttt{TWaitUnmooring}$ on every row of the file. Rao et al.\ explicitly state that they ``consider 24 (out of 35) features as independent variables, and the other features related to the calculation of total and delay time are removed,'' so the partial times are not declared as inputs. However, neither paper lists the surviving 24 features by name, and the published EDA file is the artifact a reader is most likely to reuse. A third party trying to reproduce the pipeline from the file alone, without re-reading the textual rule, could trivially include any of the partial times and recover near-perfect performance. We flag this not as a confirmed use of the partial times in the original models, but as a reproducibility hazard left open by the data release.

\subsection{Why This Counts as Target Leakage}
\label{sec:why_leakage}

Kaufman et al.~\cite{kaufman2012} classify leakage in data mining along two axes: \emph{by source} (leakage from the target itself, from the future, or from group structure) and \emph{by mechanism} (deterministic when the leaked information uniquely identifies the target, stochastic when it only correlates with it). The phenomenon we measure in §3.1 sits at the deterministic end of the target-source category: the (Mooring Port, Berth, Mooring Date) tuple is a near-key for the target, and a random row-level split places that tuple on both sides of the partition for 97\% of test berthings. This is structurally identical to the canonical example of leakage by group structure (e.g., the same patient appearing in both train and test of a medical classifier) discussed by~\cite{kaufman2012} and by recent reviews of replicability failures in applied machine learning~\cite{kapoor2023, lones2024, roelofs2019}, where the recommended fix is precisely a group-aware split. The hazard we add to that classification is that here the group-defining keys are not always explicitly present in the join output, which makes the problem easy to overlook without a deliberate audit.

\subsection{How Much of the Reported Performance is Leakage}

Table~\ref{tab:leakage_levels} already pins this down quantitatively: a Random~Forest with the same feature set as Rao et al.\ loses $0.62$ of $R^2$ (from $0.821$ to $0.204$, paired Cohen's $d = 4.24$) when the split is moved from the row level to the berthing level. Approximately that magnitude of the reported $R^2 = 0.83$ is structurally attributable to the row-level split, not to predictive content. The undocumented 202-hour ceiling separately contracts the variance of the target, which inflates the $R^2$ further on top of that. The reported Random-Forest-versus-linear gap (Table~1 of Rao et al.: $R^2 = 0.83$ for Random~Forest, $R^2 = 0.56$ for Ridge) is consistent with the same picture: tree ensembles can extract more of the row-level leakage through high-order interactions over berthing-identifying dummies than linear models can. We test this prediction explicitly in Section~\ref{sec:discussion} under the corrected pipeline.

\subsection{Pipeline Overview}
\label{sec:pipeline_overview}

\begin{figure*}[!t]
    \centering
    \includegraphics[width=\textwidth]{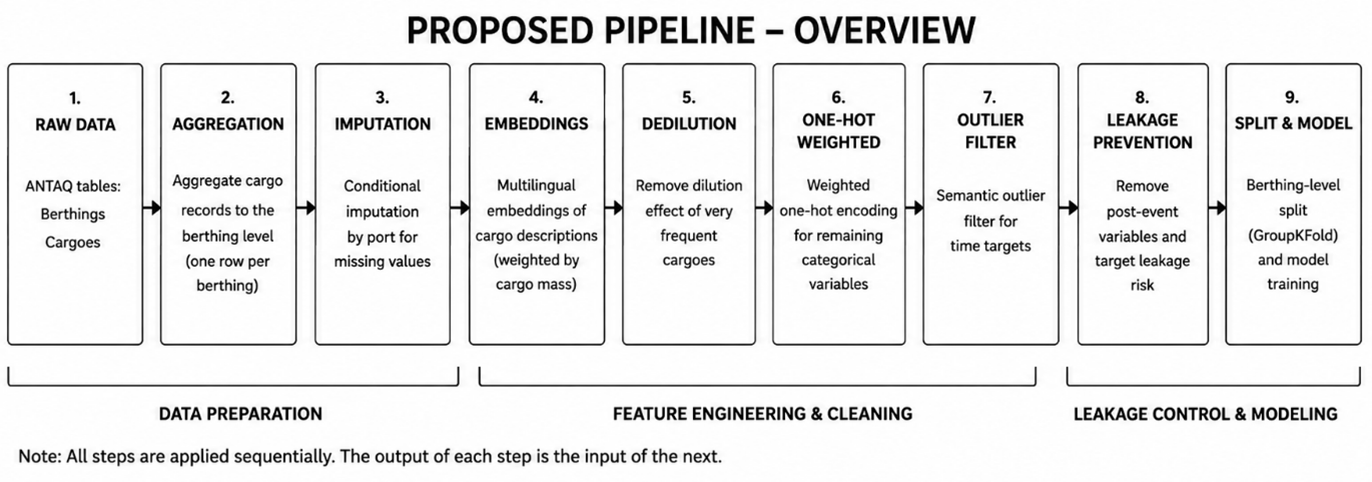}
    \caption{Overview of the proposed pipeline. The workflow comprises nine sequential stages grouped into data preparation, feature engineering and cleaning, and leakage control and modeling.}
    \label{fig:pipeline_overview}
\end{figure*}

Figure~\ref{fig:pipeline_overview} summarizes the proposed processing pipeline, from the raw ANTAQ tables to the final model training stage. The workflow is organized into nine sequential steps grouped into three main phases: data preparation, feature engineering and cleaning, and leakage control and modeling. The output of each step is used as the input to the following stage, ensuring that all transformations are applied in a consistent and reproducible order.

\section{The Proposed Pipeline}
\label{sec:pipeline}

We process the seven ANTAQ years 2018--2024 directly from the agency files. The pipeline is implemented in Python with DuckDB, Pandas, Sentence-Transformers and XGBoost and is described step by step below. After all steps the dataset contains 489{,}326 berthings split 80/20 by \texttt{IDBerthing}, with 1{,}613 features in the final matrix.

\subsection{Aggregation at the Berthing Level}

The first step rewrites the relational structure so that each row is one berthing. Numeric cargo attributes (gross weight, TEU, count) are summed; categorical attributes that are constant within a berthing (port, complex, region, navigation type, etc.) are taken with \texttt{FIRST()}; categorical attributes that vary within a berthing are expanded into a one-hot-by-cargo representation and summed with weights equal to the cargo gross mass (Section~\ref{sec:ohe_sum}); textual attributes are passed through embeddings (Section~\ref{sec:embeddings}) and aggregated with the same weights. After this step, the target appears once per berthing and a berthing-level split is well-defined.

\subsection{Conditional Imputation Instead of Row Removal}

Missing values in ANTAQ are rarely missing-completely-at-random; in the taxonomy of Little and Rubin~\cite{little2019}, most of them are missing-at-random (conditional on observable structural variables) or missing-not-at-random (the indicator of missingness itself encodes signal). The containerized cargo weight is missing whenever the cargo is not in a container; financial fields by region are missing for berthings that did not operate in that region; the route fields are missing in some terminals by construction. Removing these rows, as the previous works do, removes entire operational regimes. We instead apply three rules, motivated by the structural meaning of each ``missing'': numeric variables for which absence means non-applicability (\texttt{TEU}, \texttt{ContainerWeight}, region/river/waterway financial values) are imputed with zero; numeric variables for which absence means missingness (port coordinates, gross weight) are imputed with the median conditional on the port, with global median fallback; categorical variables of merchandise are imputed with a dedicated \emph{``Not applicable''} category so the model can learn it. More sophisticated multiple imputation schemes~\cite{vanbuuren2018} would be available, but the conditional median strategy is sufficient for the tabular boosting baseline we use and avoids introducing additional sources of variance into the ablation.

The imputer is fitted on the training set only.

\subsection{Sentence Embeddings for High-Cardinality Variables}
\label{sec:embeddings}

The four textual variables \texttt{Cargo Group}, \texttt{Containerized Cargo Group}, \texttt{Origin} and \texttt{Destination} have cardinalities of 78, 98, 342 and 466 unique values, respectively. One-hot encoding turns them into sparse 1{,}000-column blocks that hurt tree-based models with large category counts~\cite{guo2016}; dropping them, as Abreu et al.\ explicitly do under the high-cardinality criterion, loses information about routes and merchandise. A third route, originating in the word-embedding literature~\cite{mikolov2013} and adapted to tabular categorical variables by~\cite{guo2016}, is to learn a low-dimensional embedding from the training data. We choose a fourth, simpler route here: we use a pre-trained multilingual sentence encoder. The category labels in ANTAQ are not opaque identifiers; they are Portuguese phrases (e.g., ``Soja em grão'', ``Farelo de soja''), and semantically related cargoes carry semantically related labels. We encode each value once with the multilingual \texttt{paraphrase-multilingual-MiniLM-L12-v2} sentence encoder~\cite{reimers2019} (384 dimensions), which avoids the need to fit the embedding inside the predictive model. We aggregate the per-cargo embeddings into a per-berthing vector as a weighted average:
\[
E_a = \frac{\sum_i w_i \, e(c_i)}{\sum_i w_i},
\]
where $e(\cdot)$ is the embedding and $w_i$ is the gross mass of cargo $i$. A berthing dominated by one mass-heavy commodity is represented near that commodity in embedding space; a mixed berthing lies between its components weighted by mass.

Choosing embeddings does not imply that they are universally superior to categorical encoders. Both target encoding and CatBoost are effective for high-cardinality variables; however, they derive their representation from the target variable. Therefore, they must be applied separately for each target to avoid target information leakage. This requirement is relevant to the present study, as it evaluates six distinct targets and examines modeling approaches sensitive to data leakage.

Entity embeddings and learned embeddings offer a different perspective: they rely on a supervised predictive model, requiring a representation that is jointly optimized. The expected outcome depends on the model and the target making them less transparent than the four ablation configurations and they require additional setup and hyperparameter tuning.

\subsection{Dedilution of Generic Values}

One issue specific to ANTAQ is that 87.5\% of the cargo rows in 2018 list \texttt{Cargo Group} as the literal value ``Containers,'' because the specific commodity is recorded in the separate \texttt{Containerized Cargo Group} column. The weighted average over all rows therefore puts most of the mass on the generic vector for ``Containers,'' drowning the specific commodities in the mix. Symmetrically, 53\% of the cargo rows have a null \texttt{Containerized Cargo Group} because they are not containerized; if these rows go into the average for that embedding, they pull it toward ``Not applicable''. We compute each embedding restricted to the cargo rows where the value is informative: \texttt{Cargo Group} excludes the generic ``Containers,'' and \texttt{Containerized Cargo Group} excludes nulls and ``Not applicable.'' Berthings with no informative cargo for an embedding receive the zero vector, which the model learns to interpret as ``no commodity of this type.''

\subsection{Heterogeneous Categorical Cargo Variables: Weighted One-Hot-Sum}
\label{sec:ohe_sum}

Nine categorical cargo variables (e.g., \texttt{Cargo Operation Type}, \texttt{Direction}, \texttt{Container Status}, \texttt{Long Course Flag}, \texttt{Cabotage Flag}) are not constant within a berthing: between 5\% and 18\% of berthings carry more than one value, depending on the variable. A \texttt{FIRST()} aggregation here loses the mix. We instead expand each variable into one indicator per category at the cargo level, and sum each indicator weighted by gross mass at the berthing level. This produces, for each variable with $K$ categories, $K$ numeric columns that jointly encode both the dominant category and the within-berthing mix.

\subsection{Semantic Outlier Filter}

The full ANTAQ data have a long tail of berthings with extreme stay times (up to 21{,}427~h). We verified that 100\% of berthings with $\texttt{TStay} > 2{,}000$~h satisfy the conjunction \texttt{NavigationType = "Port Support" AND OperationType = "Bunkering" AND TStay > 1{,}000} h. These are support vessels in extended anchorage, typically not part of cargo planning. We remove them by a single explicit rule. The filter affects approximately 0.075\% of the dataset (369 berthings out of 489{,}695) but removes the bimodal tail that otherwise dominates the variance of the regression targets. The rule is documented and reproducible; it is not an unsupervised SVM-RBF.

\subsection{Decisions Aimed at Avoiding Leakage}
\label{sec:leak_decisions}

Three decisions are direct responses to the audit in Section~\ref{sec:critique}: \texttt{Berth} is excluded because it is assigned after arrival (the prediction would not be useful if it required knowing where the ship will actually moor); absolute dates are excluded because their pairwise differences reconstruct the target; the partial times are kept in the dataset only as alternative regression targets, never as inputs. The target encoding of \texttt{Mooring Port} follows the procedure of Micci-Barreca~\cite{miccibarreca2001} with a smoothing fallback to the global median for ports unseen in training, and is fitted only on the inner training core (excluding the early-stopping hold-out) to avoid the leakage that target encoding introduces if applied across the full training set.

\subsection{Stratified Split by Berthing}

The 80/20 split is performed over \texttt{IDBerthing}, stratified by Geographic Region, with an assertion of disjoint training and test berthings. This is the most direct response to the leakage measured in Section~\ref{sec:critique}.

\subsection{Geographic and Cyclic Features}

Continuous longitude and latitude of the port are included in the baseline, complementing the categorical geographical variables: tree splits can use them to detect proximity beyond exact matches on \texttt{Port} or \texttt{Region}. The month is encoded both as an integer and as $(\sin 2\pi m/12, \cos 2\pi m/12)$ so December and January are near each other in the cyclic representation.

\subsection{Final Feature Matrix}
\label{sec:features}

After the operations above, the M4 configuration uses 1{,}606 features per berthing as model inputs. Table~\ref{tab:feature_breakdown} summarizes their composition. \textbf{Berthing-level descriptors} (13 categoricals and 7 numerics, including longitude/latitude and the cyclic encoding of the month) identify the operating context of the port and the date. \textbf{Cargo-derived numerics} (17 columns) carry the total mass, count, container variables, and the financial values aggregated by hydrography. Three \textbf{distinct-value counters} (\texttt{qt\_distinct\_*}) flag berthings whose cargo records carry more than one value of a near-homogeneous categorical variable. \textbf{Cargo categorical mixtures} (32 OHE-sum columns, Section~\ref{sec:ohe_sum}) carry the within-berthing distribution of operation types, container statuses and flags. The remaining 1{,}536 columns are the four 384-dimensional embedding blocks (origin, destination, cargo group, containerized cargo group); origin and destination embeddings are always in the baseline, while the two cargo embeddings are the variables under ablation (M1: none; M2: cargo group only; M3: containerized only; M4: both). The end-to-end flow can be read as a sequence of seven steps applied to the source tables of Section~\ref{sec:data}: (1) load the seven annual ANTAQ tables and join the cargo satellites onto \texttt{Carga}; (2) encode the four text variables once with the multilingual sentence model; (3) aggregate cargo to the berthing level---summing numerics, taking \texttt{FIRST()} on homogeneous categoricals, weighted one-hot-sum on heterogeneous ones, and the mass-weighted, dediluted embedding average on the text blocks; (4) apply the semantic outlier filter; (5) split 80/20 by \texttt{IDBerthing}, stratified by region; (6) impute missing values with the train-only conditional rules and target-encode the mooring port on the inner training core; and (7) feed the resulting matrix to the four-configuration ablation.

\begin{table}[!t]
\centering
\caption{Composition of the M4 feature set (1{,}606 model inputs). M1 omits the last two rows; M2 omits the last row; M3 omits the second-to-last row.}
\label{tab:feature_breakdown}
\small
\setlength{\tabcolsep}{4pt}
\begin{tabular}{@{}p{0.78\columnwidth}r@{}}
\toprule
\textbf{Block} & \textbf{Columns} \\ \midrule
Berthing-level categoricals (port,  UF, region, \dots) & 13 \\
Berthing-level numerics (Year, Month, DoW, $\sin/\cos$ of month, lon, lat) & 7 \\
Cargo-derived numerics (gross weight, TEU, count, financial \dots) & 17 \\
Distinct-value counters \texttt{qt\_distinct\_*} & 3 \\
Cargo categorical mixtures (OHE-sum, 9 variables) & 32 \\
\texttt{Origin} embedding (384 dims) & 384 \\
\texttt{Destination} embedding (384 dims) & 384 \\
\texttt{Cargo Group} embedding (M2, M4) & 384 \\
\texttt{Containerized Cargo Group} embedding (M3, M4) & 384 \\
\midrule
\textbf{Total inputs to the M4 model} & \textbf{1{,}606} \\
\bottomrule
\end{tabular}
\end{table}

\section{Experimental Design}
\label{sec:experiment}

\begin{figure}[!t]
\centering
\includegraphics[width=\columnwidth]{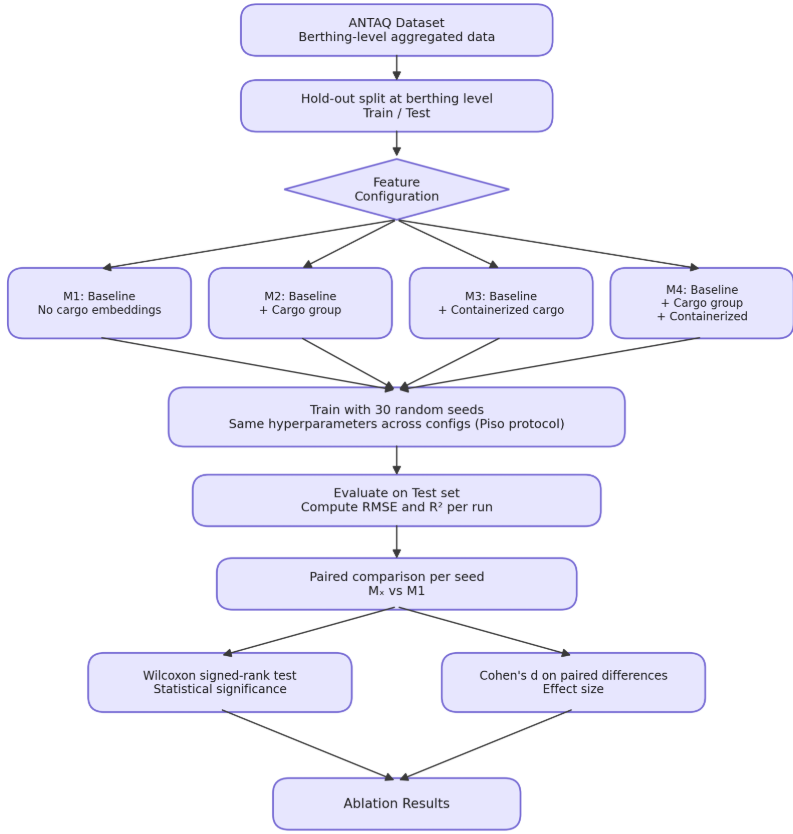}
\caption{The four configurations of the ablation share an identical baseline (M1) and differ only in which cargo embeddings are added.}
\label{fig:ablation}
\end{figure}

We use the seven ANTAQ years 2018--2024, after the pipeline of Section~\ref{sec:pipeline} produces 489{,}326 berthings (391{,}460 train, 97{,}866 test). We predict six time targets in hours: \texttt{TStay}, \texttt{TWaitMooring}, \texttt{TMoored}, \texttt{TWaitStartOp}, \texttt{TOperation}, \texttt{TWaitUnmooring}. We evaluate four configurations that differ only in whether cargo embeddings are included (Fig.~\ref{fig:ablation}). M1 is the baseline with all features described in Section~\ref{sec:pipeline} except the cargo embeddings (833 features); M2 adds the 384 dimensions of the \texttt{Cargo Group} embedding (1{,}222 features); M3 adds the 384 dimensions of the \texttt{Containerized Cargo Group} embedding (1{,}222 features); M4 adds both (1{,}606 features). Quantitative containerized variables (TEU, count, total containerized weight) remain in the baseline, so that the variable under ablation is the \emph{semantic} content and not the quantity.
 XGBoost~\cite{chen2016} hyperparameters are tuned once per target using Optuna with TPE for 50 trials on M1 and reused across M2, M3 and M4 for that target. This \emph{ceteris paribus} protocol is conservative: a configuration that would prefer different hyperparameters pays a price. Hyperparameters differ \emph{across} targets (the time scales differ by more than an order of magnitude) but never \emph{across} configurations within the same target. We train 30 seeds per (target $\times$ configuration), for a total of 720 models. For each pairwise comparison $\text{M}_x$ vs M1 we compute the paired difference of RMSE over the 30 seeds and report the mean difference, Cohen's $d$ on the paired differences~\cite{cohen1988}, and the one-sided paired Wilcoxon $p$-value~\cite{wilcoxon1945} for ``M1 worse than $\text{M}_x$.'' Cohen's conventions~\cite{cohen1988} label $|d| = 0.2$ as small, $0.5$ as medium and $0.8$ as large, and the extension by Sawilowsky~\cite{sawilowsky2009} adds $1.2$ as very large, $2.0$ as huge and $3.0$ as effectively deterministic. In a paired design with 30 seeds, $|d| \geq 0.5$ is already detectable at $p<0.05$.

XGBoost was selected given the dataset's characteristics: it consists of tabular data with numerical variables, categorical variables, weighted load representations, and target encoding. As a gradient-boosted tree model, it is well-suited for capturing non-linear relationships. The study does not aim to compare all algorithm families but rather to help understand semantic load representations within a reproducible, leakage-free pipeline. Therefore, XGBoost is retained across all ablation configurations to confirm performance differences based on the presence or absence of the embeddings.

Using LightGBM, CatBoost, and neural networks offers an interesting alternative for future comparisons. LightGBM may provide computational advantages, while CatBoost is suitable for categorical variables though in the current scenario, these are already represented via embeddings. Neural networks, meanwhile, would require additional measures such as regularization and fine-tuning; this would expand the scope and complicate the isolation of semantic load representations, which is the objective of this analysis.

Direct comparison of absolute RMSE and $R^2$ between the prior work and ours is not informative: the populations differ. The published files have an undocumented 202-h ceiling and retain partial-time variables that algebraically sum to the target; our pipeline preserves the realistic distribution after the explicit semantic filter and excludes the partial times from the inputs. We therefore report absolute metrics on our pipeline and use the prior work only as a description of the methodology that produced the inflated numbers.

\section{Results}
\label{sec:results}

Table~\ref{tab:abs_metrics} reports the 30-seed means; Tables~\ref{tab:effects} and~\ref{tab:effects_r2} report the paired effects of cargo embeddings on RMSE and on $R^2$, respectively. Because $R^2$ is a monotone transform of RMSE on a fixed test set, the two effect tables agree in sign and significance; we report both. 

\begin{table}[!t]
\centering
\caption{Absolute performance on the test set, 30 seeds per cell. Mean $\pm$ standard deviation.}
\label{tab:abs_metrics}
\small
\setlength{\tabcolsep}{4pt}
\resizebox{\columnwidth}{!}{%
\begin{tabular}{lllll}
\toprule
\textbf{Target} & \textbf{M1} & \textbf{M2} & \textbf{M3} & \textbf{M4} \\
\midrule
\multicolumn{5}{l}{\textit{RMSE (hours)}} \\
TStay              & 76.21$\,\pm\,$0.10 & 74.88$\,\pm\,$0.08 & 76.24$\,\pm\,$0.09 & 74.87$\,\pm\,$0.08 \\
TWaitMooring       & 67.10$\,\pm\,$0.07 & 65.85$\,\pm\,$0.06 & 67.11$\,\pm\,$0.07 & 65.85$\,\pm\,$0.06 \\
TMoored            & 28.97$\,\pm\,$0.06 & 28.66$\,\pm\,$0.05 & 28.96$\,\pm\,$0.06 & 28.64$\,\pm\,$0.05 \\
TWaitStartOp       & 18.41$\,\pm\,$0.04 & 18.45$\,\pm\,$0.04 & 18.43$\,\pm\,$0.05 & 18.46$\,\pm\,$0.06 \\
TOperation         & 19.34$\,\pm\,$0.02 & 19.08$\,\pm\,$0.02 & 19.35$\,\pm\,$0.04 & 19.08$\,\pm\,$0.03 \\
TWaitUnmooring     &  6.43$\,\pm\,$0.02 &  6.43$\,\pm\,$0.01 &  6.43$\,\pm\,$0.01 &  6.42$\,\pm\,$0.01 \\
\midrule
\multicolumn{5}{l}{\textit{$R^2$}} \\
TStay              & 0.6195 & 0.6327 & 0.6192 & 0.6328 \\
TWaitMooring       & 0.5617 & 0.5778 & 0.5615 & 0.5778 \\
TMoored            & 0.5936 & 0.6022 & 0.5938 & 0.6027 \\
TWaitStartOp       & 0.2874 & 0.2841 & 0.2860 & 0.2830 \\
TOperation         & 0.6989 & 0.7069 & 0.6985 & 0.7069 \\
TWaitUnmooring     & 0.3598 & 0.3597 & 0.3585 & 0.3601 \\
\bottomrule
\end{tabular}%
}
\end{table}

\begin{table}[!t]
\centering
\caption{Paired effect of cargo embeddings vs.\ M1 (30 seeds). $\Delta\mathrm{RMSE}$ is the mean of M1$-$M$_x$ per seed (positive means M$_x$ better). $d$ is Cohen's $d$ on the paired differences. $p$ is the one-sided paired Wilcoxon. Bold entries are significant at $p<0.05$.}
\label{tab:effects}
\small
\setlength{\tabcolsep}{4pt}
\resizebox{\columnwidth}{!}{%
\begin{tabular}{llrrr}
\toprule
\textbf{Target} & \textbf{Comp.} & $\boldsymbol{\Delta\mathrm{RMSE}}$ \textbf{(h)} & \textbf{Cohen's $d$} & \textbf{$p$} \\
\midrule
\multirow{3}{*}{TStay}
 & M2 vs M1 & \textbf{+1.328} & \textbf{11.83} & \textbf{$<10^{-7}$} \\
 & M3 vs M1 & $-$0.031        & $-$0.24        & 0.897 \\
 & M4 vs M1 & \textbf{+1.342} & \textbf{10.17} & \textbf{$<10^{-7}$} \\
\midrule
\multirow{3}{*}{TWaitMooring}
 & M2 vs M1 & \textbf{+1.249} & \textbf{15.41} & \textbf{$<10^{-7}$} \\
 & M3 vs M1 & $-$0.016        & $-$0.16        & 0.809 \\
 & M4 vs M1 & \textbf{+1.245} & \textbf{17.28} & \textbf{$<10^{-7}$} \\
\midrule
\multirow{3}{*}{TMoored}
 & M2 vs M1 & \textbf{+0.306} & \textbf{3.76}  & \textbf{$<10^{-7}$} \\
 & M3 vs M1 & +0.005          & 0.06           & 0.271 \\
 & M4 vs M1 & \textbf{+0.326} & \textbf{4.69}  & \textbf{$<10^{-7}$} \\
\midrule
\multirow{3}{*}{TWaitStartOp}
 & M2 vs M1 & $-$0.042        & $-$0.80        & 1.000 \\
 & M3 vs M1 & $-$0.018        & $-$0.32        & 0.940 \\
 & M4 vs M1 & $-$0.056        & $-$0.98        & 1.000 \\
\midrule
\multirow{3}{*}{TOperation}
 & M2 vs M1 & \textbf{+0.259} & \textbf{7.44}  & \textbf{$<10^{-7}$} \\
 & M3 vs M1 & $-$0.013        & $-$0.33        & 0.971 \\
 & M4 vs M1 & \textbf{+0.259} & \textbf{5.62}  & \textbf{$<10^{-7}$} \\
\midrule
\multirow{3}{*}{TWaitUnmooring}
 & M2 vs M1 & $-$0.000        & $-$0.01        & 0.556 \\
 & M3 vs M1 & $-$0.006        & $-$0.28        & 0.897 \\
 & M4 vs M1 & +0.002          & 0.07           & 0.159 \\
\bottomrule
\end{tabular}%
}
\end{table}

\begin{table}[!t]
\centering
\caption{Paired effect of cargo embeddings vs.\ M1 on $R^2$ (30 seeds). $\Delta R^2$ is the mean of M$_x-$M1 per seed (positive means M$_x$ better). $d$ is Cohen's $d$ on the paired differences. $p$ is the one-sided paired Wilcoxon. Bold entries are significant at $p<0.05$.}
\label{tab:effects_r2}
\small
\setlength{\tabcolsep}{4pt}
\resizebox{\columnwidth}{!}{%
\begin{tabular}{llrrr}
\toprule
\textbf{Target} & \textbf{Comp.} & $\boldsymbol{\Delta R^2}$ & \textbf{Cohen's $d$} & \textbf{$p$} \\
\midrule
\multirow{3}{*}{TStay}
 & M2 vs M1 & \textbf{+0.0132} & \textbf{11.80} & \textbf{$<10^{-7}$} \\
 & M3 vs M1 & $-$0.0003        & $-$0.24        & 0.897 \\
 & M4 vs M1 & \textbf{+0.0133} & \textbf{10.16} & \textbf{$<10^{-7}$} \\
\midrule
\multirow{3}{*}{TWaitMooring}
 & M2 vs M1 & \textbf{+0.0162} & \textbf{15.37} & \textbf{$<10^{-7}$} \\
 & M3 vs M1 & $-$0.0002        & $-$0.16        & 0.809 \\
 & M4 vs M1 & \textbf{+0.0161} & \textbf{17.21} & \textbf{$<10^{-7}$} \\
\midrule
\multirow{3}{*}{TMoored}
 & M2 vs M1 & \textbf{+0.0086} & \textbf{3.76}  & \textbf{$<10^{-7}$} \\
 & M3 vs M1 & +0.0002          & 0.06           & 0.271 \\
 & M4 vs M1 & \textbf{+0.0091} & \textbf{4.69}  & \textbf{$<10^{-7}$} \\
\midrule
\multirow{3}{*}{TWaitStartOp}
 & M2 vs M1 & $-$0.0032        & $-$0.80        & 1.000 \\
 & M3 vs M1 & $-$0.0014        & $-$0.32        & 0.940 \\
 & M4 vs M1 & $-$0.0044        & $-$0.98        & 1.000 \\
\midrule
\multirow{3}{*}{TOperation}
 & M2 vs M1 & \textbf{+0.0080} & \textbf{7.43}  & \textbf{$<10^{-7}$} \\
 & M3 vs M1 & $-$0.0004        & $-$0.33        & 0.971 \\
 & M4 vs M1 & \textbf{+0.0080} & \textbf{5.63}  & \textbf{$<10^{-7}$} \\
\midrule
\multirow{3}{*}{TWaitUnmooring}
 & M2 vs M1 & $-$0.0000        & $-$0.01        & 0.548 \\
 & M3 vs M1 & $-$0.0013        & $-$0.28        & 0.897 \\
 & M4 vs M1 & +0.0004          & 0.07           & 0.159 \\
\bottomrule
\end{tabular}%
}
\end{table}

\begin{figure}[!t]
\centering
\includegraphics[width=\columnwidth]{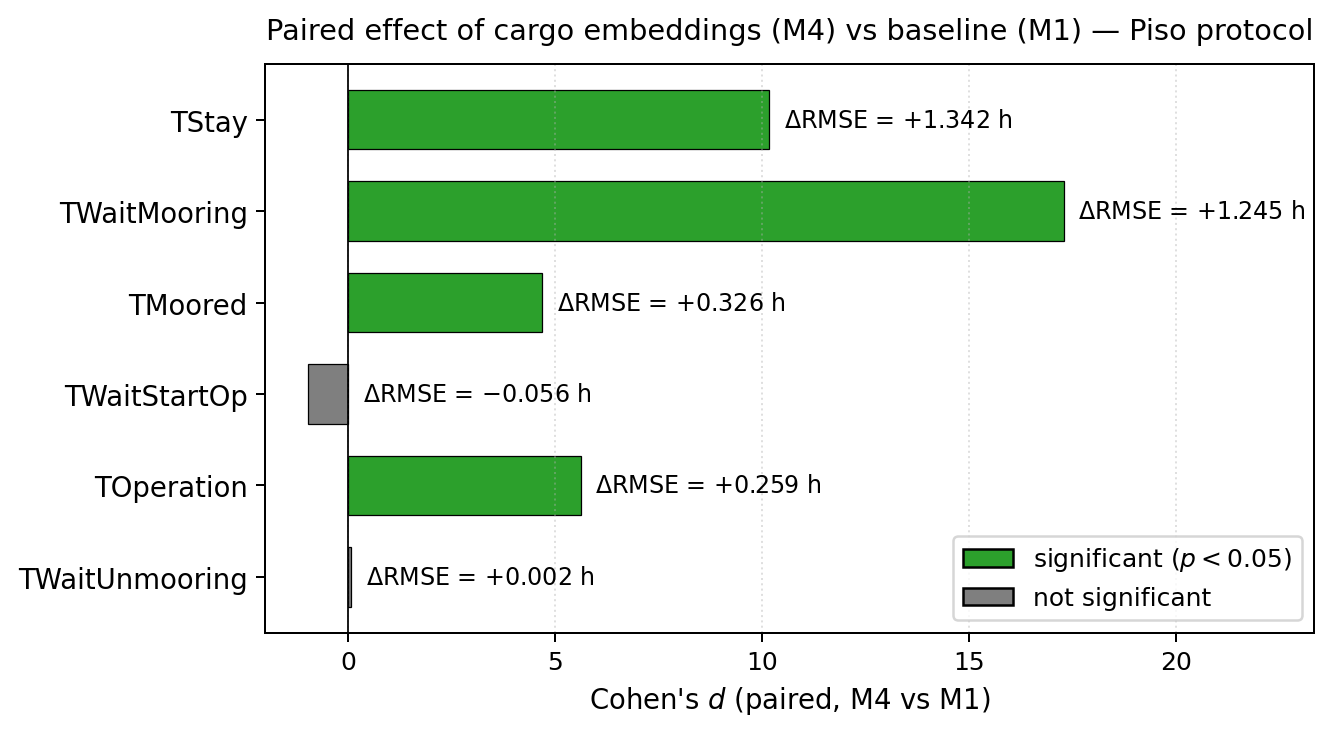}
\caption{Paired effect of M4 (both cargo embeddings) vs.\ M1 (baseline) by target. Bars show Cohen's $d$ on the paired RMSE differences over 30 seeds; the label at the end of each bar shows $\Delta$RMSE in hours. Green bars are significant at $p<0.05$ (paired Wilcoxon, one-sided); grey bars are not.}
\label{fig:effects}
\end{figure}

Four out of six targets carry a clear cargo signal. Adding the \texttt{Cargo Group} embedding (M2) yields a statistically robust improvement on \texttt{TStay} ($\Delta\mathrm{RMSE} = +1.33$~h, $d = 11.83$, $p < 10^{-7}$), \texttt{TWaitMooring} ($+1.25$~h, $d = 15.41$), \texttt{TMoored} ($+0.31$~h, $d = 3.76$) and \texttt{TOperation} ($+0.26$~h, $d = 7.44$). The absolute reductions track the scale of each target, but the effect sizes consistently fall above the conventional threshold for a large effect: values of $d$ above $3$ correspond to the embedding winning on essentially every seed. M4, which adds both embeddings, is statistically indistinguishable from M2 on these targets, indicating that the \texttt{Containerized Cargo Group} embedding contributes little marginal information once the broader cargo group is in.

M3 alone is consistently non-significant. The containerized cargo embedding by itself never beats M1 at $p < 0.05$ on any of the six targets. The quantitative containerized variables already in the baseline (TEU, count, total weight) absorb the available signal; adding the semantic vector of the containerized commodity alone does not improve M1, contrary to the intuition that container information would be the main driver of port operations. The remaining two targets are insensitive to cargo information. \texttt{TWaitStartOp} and \texttt{TWaitUnmooring}---the administrative waiting times after the vessel is already moored---show no positive effect from adding cargo embeddings (\texttt{TWaitStartOp} exhibits a small negative effect of approximately $-0.05$~h, $|d| < 1$). These intervals are governed by administrative procedures rather than by what is being loaded, so the absence of effect is the expected outcome. The prior literature aggregates these intervals into a single stay-time target and therefore cannot resolve the heterogeneity.

The absolute RMSE of \texttt{TStay} M1 (76.21~h) is not directly comparable with the 10.44~h reported by Rao et al. Their figure is on a population truncated at 202~h with the partial-time variables available as inputs, and on a 20\% test set heavily overlapped with the training set under the row-level split. Our 76.21~h is on a population that includes the realistic tail of long berthings up to 1{,}000~h with strict berthing-level separation between training and test. The two numbers measure different quantities and any comparison should be made on the paired effect sizes within a single pipeline, not on absolute RMSE across pipelines.

The figure 6 helps to compare our work with the international context. It can be observed that higher R² values appear in the studies reviewed; however, this is due to differences in target variables, data sources, and operational contexts.

\begin{figure*}[!t]
    \centering
    \includegraphics[width=0.96\textwidth]{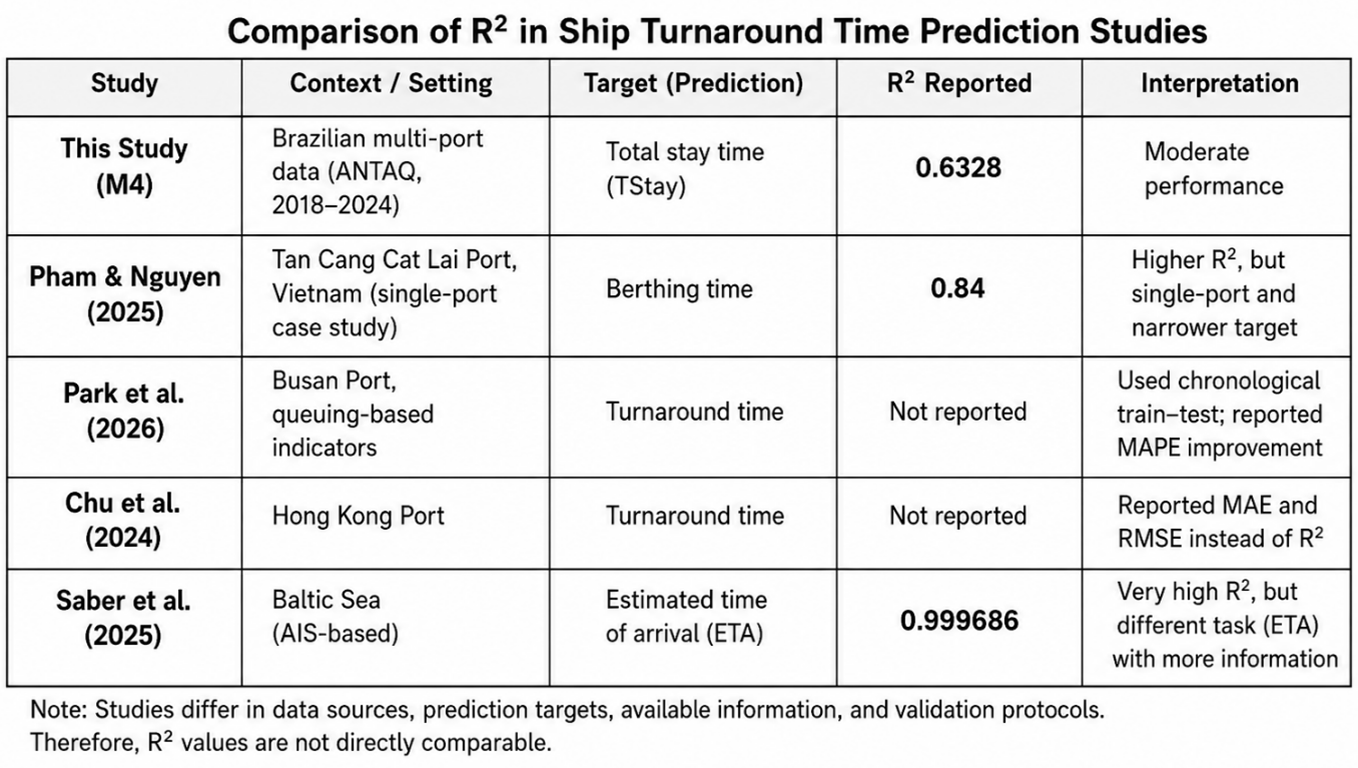}
    \caption{Comparison of reported $R^2$ values in studies on ship
    turnaround-time prediction. The studies differ in their prediction
    targets, data sources, operational contexts, available information,
    and validation protocols; therefore, the reported values are not
    directly comparable.}
    \label{fig:international_comparison}
\end{figure*}

\section{Discussion}
\label{sec:discussion}

Rao et al.\ report Random~Forest at $R^2 = 0.83$ and Ridge at $R^2 = 0.56$ and attribute the gap to the tree's ability to handle categorical interactions. The Random~Forest row of Table~\ref{tab:leakage_levels} reproduces the upper end of that gap ($0.821$) under the row-level split; under the berthing-level split the same Random~Forest collapses to $0.204$, while a Ridge regressor on the same data exhibits a smaller drop. The portion of the tree-versus-linear gap that survives the corrected split is much narrower than the published numbers suggest. Tree ensembles can extract more of the row-level identity leakage through high-order interactions over target-encoded berthing dummies than linear models can; once the identity is no longer leaked, that advantage diminishes. Once the six time components are separated and analyzed independently, cargo identity matters substantially for the operational components ($d \approx 17$ on \texttt{TWaitMooring}, $\approx 10$ on \texttt{TStay}, between $4$ and $7$ on \texttt{TMoored} and \texttt{TOperation}) and is not measurable for the administrative ones (waits to start and finish the paperwork). A single aggregate stay-time target averages large per-cargo effects on operational intervals with zero per-cargo effect on administrative ones; the aggregate effect is real but not actionable for an operator who needs to know which time component is sensitive to cargo. The pipeline does not currently incorporate weather, queue length at the destination port, or vessel-side telemetry. The outlier filter is a single explicit conjunction; a more refined version could partition the very-long-stay regime (port-support bunkering) into its own model rather than dropping it.

The results show that cargo embeddings are specific to each target metric, rather than being uniform across the vessel's stay process. Cargo information helps improve predictions for TStay, TWaitMooring, TMoored, and TOperation, though it is less effective for TWaitStartOp or TWaitUnmooring. This pattern is operationally ideal. Characteristics such as volume, destination, and handling requirements are linked to the demand for berths, loading and unloading operations, equipment, and the time required to process the cargo. These representations provide information not captured by aggregated variables like total weight, TEU, or container count.

The lack of improvement regarding administrative waiting targets should not be interpreted to mean that this data is unimportant; rather, it indicates that the available cargo variable does not provide sufficient information about that specific phase. The variables TWaitStartOp and TWaitUnmooring depend more on documentation readiness, terminal coordination, and clearance procedures, as they relate to the vessel once it is already berthed. This data is not directly reflected in the cargo descriptions. Thus, the discrepancy between operational and administrative targets arises because cargo characteristics have a stronger impact on operations than on administrative activities following berthing. This interpretation is associative rather than causal.

One consequence is that a single residence time target can mask heterogeneity. Predicting only residence time shows that information regarding load is useful, yet it does not reveal which part of the residence process is affected. The distinction made allows for an actionable interpretation: models that account for load are better at predicting workload and delays, whereas the administrative aspect requires additional variables.

For managers, cargo composition should be incorporated into the physical operations planning tool. Semantic representations of the cargo can facilitate more informed predictions regarding waiting time, time at berth, handling time, and total turnaround time. These predictions support berth allocation management, while also aiding in supply and supply chain management.

 However, generalizing the cargo-based model into a universal solution for addressing delays is not ideal. The analysis shows that cargo related metrics do not capture administrative aspects; therefore, managers seeking to reduce timeframes should utilize operational indicators currently missing from the model such as document processing status, inspection requirements, and coordination among port agents. Thus, the analysis supports a holistic management approach: while cargo data is useful for physical planning, managing administrative performance requires a new approach to process management.

The main experiment uses a \emph{floor} protocol that tunes hyperparameters once on M1 per target and reuses them across M2, M3 and M4 under the \emph{ceteris paribus} rule of Section~\ref{sec:experiment}. As an adversarial control we also ran a \emph{ceiling} variant that re-tunes hyperparameters independently for each of the 24 (target, configuration) cells with the same Optuna budget (50 trials $\times$ 5-fold CV). All other elements are identical: same train/test split, same \texttt{train\_test\_split} for early-stopping validation, same hold-out evaluation, same 30 seeds. Table~\ref{tab:ceiling_rmse} reports the Ceiling absolute performance. Across the 18 cells where the two protocols are not identical by construction (M2, M3, M4), the absolute paired difference Ceiling $-$ Floor averages $0.055$~h, the cell-wise mean difference is bounded by $|\Delta| \leq 0.17$~h, and the Floor protocol outperforms the Ceiling protocol in 11 of those 18 cells. This pattern is consistent with selection bias from picking the best of 50 noisy cross-validation estimates per cell~\cite{cawley2010}: hyperparameters tuned per cell win their internal cross-validation but pay an optimistic-bias tax on hold-out, while M1's hyperparameters transferred under the Floor protocol are statistically independent of M2/M3/M4 fold noise and act as an implicit regularizer. The bias is largest on the targets with no genuine cargo signal (\texttt{TWaitStartOp}: Ceiling $0.17$~h worse on M4), where the search exploits fold-specific noise that does not generalize. The M2- and M4-vs-M1 effects are therefore robust to whether hyperparameters are transferred or re-tuned per cell. A repeated $K$-fold tuning objective, or aligning the tuning protocol to the early-stopping validation split used at training time, would reduce the bias further.

\begin{table}[!t]
\centering
\caption{Absolute performance under the Ceiling protocol (hyperparameters re-tuned per cell), 30 seeds per cell. Mean $\pm$ standard deviation. Compare against Floor results in Table~\ref{tab:abs_metrics}.}
\label{tab:ceiling_rmse}
\small
\setlength{\tabcolsep}{4pt}
\resizebox{\columnwidth}{!}{%
\begin{tabular}{lllll}
\toprule
\textbf{Target} & \textbf{M1} & \textbf{M2} & \textbf{M3} & \textbf{M4} \\
\midrule
\multicolumn{5}{l}{\textit{RMSE (hours)}} \\
TStay              & 76.21$\,\pm\,$0.10 & 74.98$\,\pm\,$0.11 & 76.26$\,\pm\,$0.08 & 74.91$\,\pm\,$0.09 \\
TWaitMooring       & 67.10$\,\pm\,$0.07 & 65.83$\,\pm\,$0.06 & 67.03$\,\pm\,$0.07 & 65.74$\,\pm\,$0.06 \\
TMoored            & 28.97$\,\pm\,$0.06 & 28.66$\,\pm\,$0.07 & 29.03$\,\pm\,$0.06 & 28.68$\,\pm\,$0.07 \\
TWaitStartOp       & 18.41$\,\pm\,$0.04 & 18.43$\,\pm\,$0.09 & 18.52$\,\pm\,$0.06 & 18.63$\,\pm\,$0.05 \\
TOperation         & 19.34$\,\pm\,$0.02 & 19.08$\,\pm\,$0.02 & 19.36$\,\pm\,$0.02 & 19.10$\,\pm\,$0.02 \\
TWaitUnmooring     &  6.43$\,\pm\,$0.02 &  6.45$\,\pm\,$0.02 &  6.42$\,\pm\,$0.01 &  6.42$\,\pm\,$0.01 \\
\midrule
\multicolumn{5}{l}{\textit{$R^2$}} \\
TStay              & 0.6195 & 0.6317 & 0.6191 & 0.6324 \\
TWaitMooring       & 0.5617 & 0.5780 & 0.5626 & 0.5792 \\
TMoored            & 0.5936 & 0.6023 & 0.5919 & 0.6018 \\
TWaitStartOp       & 0.2874 & 0.2854 & 0.2785 & 0.2699 \\
TOperation         & 0.6989 & 0.7068 & 0.6981 & 0.7063 \\
TWaitUnmooring     & 0.3598 & 0.3558 & 0.3601 & 0.3608 \\
\bottomrule
\end{tabular}%
}
\end{table}

All code, the cleaned 2018--2024 dataset, the hyperparameter files, and the 720 seed-by-seed model artifacts are released. The pipeline runs end to end on the raw ANTAQ files. The empirical audit of Section~\ref{sec:critique} is shipped as an executable notebook that processes the published CSVs directly and produces every number quoted in this paper.

\section{Threats to Validity}
\label{sec:threats_validity}

\subsection{Internal Validity}

\textbf{Dependency on ANTAQ:} Our empirical audit and the proposed reproducible pipeline heavily rely on the data provided by the Brazilian National Waterway Transportation Agency (ANTAQ). Any systemic reporting errors, hidden biases, or structural artifacts inherent to the ANTAQ database schema could directly influence our findings regarding data leakage and predictive performance.

\subsection{External Validity}

\textbf{Applicability to other countries:} The structural leakage from one-to-many joins and the textual representations we evaluated are specifically tied to Brazilian port operations and the Portuguese labels within the ANTAQ dataset. Consequently, it remains uncertain whether these specific methodological pitfalls and the performance gains from semantic embeddings fully generalize to seaport operations and maritime data structures in other countries.

\subsection{Construct Validity}

\textbf{Choice of targets:} To model the ship turnaround time, we decomposed the total stay into six separate operational and administrative time targets (\texttt{TStay}, \texttt{TWaitMooring}, \texttt{TMoored}, \texttt{TWaitStartOp}, \texttt{TOperation}, and \texttt{TWaitUnmooring}) treated as distinct regression problems. While this choice successfully reveals heterogeneous, target-specific effects of cargo representation, it relies on the assumption that these specific six intervals adequately and comprehensively capture the construct of port operational efficiency.

\subsection{Statistical Validity}

\textbf{Number of seeds:} To ensure the reliability of our results and account for variance, our controlled ablation study evaluated the predictive models over 30 random seeds. Although 30 seeds provide a robust sample to estimate model stability, evaluating a different finite number of seeds could marginally alter the variance and confidence intervals of the estimated metrics.

\textbf{Statistical tests used:} We utilized paired Wilcoxon tests to evaluate statistical significance and calculated Cohen's $d$ to measure the effect size of our performance improvements. While these non-parametric tests are highly appropriate for our experimental design, the application of alternative statistical frameworks could potentially yield different interpretations regarding the magnitude of the semantic embeddings' impact.

\section{Conclusion}

The headline numbers in the prior published work on ship turnaround time at Brazilian ports were produced by a pipeline that combines a one-to-many join with a row-level split and an undocumented truncation of the target distribution. The combination is enough to drive a trivial lookup that ignores all features to $R^2 = 0.99$ on the published split. The corrected pipeline we propose produces performance estimates that we believe are more representative of the underlying problem, and it makes visible a target-specific structure that a single-target analysis does not capture: cargo identity has very large and statistically robust effect on the operational components of the stay (loading and the wait to moor) and no measurable effect on the administrative ones. We do not claim that the prior authors made any computational mistake; we claim that the combination of data-release and methodology choices that they describe, together with the structure of the source database, produces numbers whose interpretation as predictive performance is misleading.

\vfill
\end{document}